\documentstyle[prl,aps]{revtex}
\tightenlines
\twocolumn
\begin{document}
\draft
\title{Universal relation between longitudinal and transverse\\
conductivities in quantum Hall effect}
\author{Igor Ruzin$^1$ and Shechao Feng$^{1,2}$}
\address{1. Department of Physics, Unversity of California,
Los Angeles, CA 90024-1547, USA;\\
2. Department
of Physics, Hong Kong University of Science and
           Technology,\\
 Clear Water Bay Road, Kowloon, Hong Kong }
\date{ \today }
\maketitle
\begin{abstract}
We show that any critical transition region
between two adjacent Hall plateaus in either
integer or fractional quantum Hall effect
is characterized by a universal ``semi-circle''
relationship between the longitudinal and transverse conductivities
$\sigma_{xx}$,
$\sigma_{xy}$, provided the sample is homogeneous and isotropic on a large
scale.
This is demonstrated both for the phase-coherent quantum transport as well as
for the
incoherent transport.

\end{abstract}

\vskip 1truecm

\narrowtext

Recently there has been much interest in the study of transition
regions between successive quantum Hall plateaus, in both the
integer and the fractional regime.
Most of these studies have been focused on
how the transition region width $\Delta B$ scales with temperature
at low temperatures.
Experimentally, it has been noted that the critical dependence $\Delta B
\propto T^\kappa$
can be extracted either directly
from the half-width of the peak in $\sigma_{xx} (B)$, or
from the sharpness of the step in Hall conductivity
$\sigma_{xy}(B)$.
Thus it has been realized, on intuitive ground at least, that the peak-like
$\sigma_{xx} (B)$ function and the step-like $\sigma_{xy}(B)$ function in the
transition region between two successive plateaus are related.  Based
on a semiclassical model that the transition region, in the critical regime, is
represented by a
random mixture of two liquids with different quantized local Hall
conductivities
$\sigma_1$ and $\sigma_2$, Dykhne and Ruzin\cite{dykhne}
recently developed a phenomenological theory to show that
$\sigma_{xx}$ and $\sigma_{xy}$ are related by a simple
``semi-circle'' relation:
\begin{equation}
\sigma_{xx}^2 +\left(\sigma_{xy}-\frac{\sigma_1+\sigma_2}{2} \right)^2
=\left(\frac{\sigma_1-\sigma_2}{2} \right)^2 .
\label{semi circle}
\end{equation}
Physically, a finite {\it effective} longitudinal conductivity $\sigma_{xx}$ at
$T\to 0$ is related to the existence of singularities in the current
distribution (``hot spots")
close to the percolation threshold.
To derive relation~(\ref{semi circle}), authors in Ref. \cite{dykhne} assumed a
very weak
scattering in the bulk and described it by a {\it local} diagonal
conductivity $\sigma_{xx}^{{\rm loc}} \ll \sigma _2-\sigma _1$, defined on a
scale shorter than the
correlation length of the two-phase mixture. This assumption is essentially
classical and does not apply to the more experimentally relevant
low-temperature regime
where quantum coherence exists on a much larger scale\cite{dykhne}.

In this letter, we demonstrate that the same simple relationship between
$\sigma_{xx}$ and $\sigma_{xy}$ holds in the  quantum coherent transport regime
as well.  We achieve this by first proving a theorem
that the semi-circle relation between $\sigma_{xx}$ and $\sigma_{xy}$ is
equivalent
to the perpendicularity between the average current densities in the two
phases.
We then show by simulations on the quantum percolation model
that the currents are indeed perpendicular to a high accuracy, thus
finishing the demonstration.

As in Ref. \cite{dykhne}, let us consider two competing quantum Hall liquids in
the presence of a
long-range random potential, with the magnetic field determining volume
fractions (close to $1/2$), as
depicted in Fig.~\ref{quantum perco}(a). The ``white'' regions represent the
phase with
(quantized) Hall conductivity $\sigma_1$, such that if this phase percolates
freely throughout the
sample, the system will be on the Hall plateau with $\sigma_{xy} = \sigma_1$.
Similarly, the
``black'' regions have local Hall conductivity $\sigma_2$.  Near the
percolation threshold, transport
is controlled by quantum tunneling through the various saddle points
(vertices), one of which is
shown in detail in Fig.~\ref{quantum perco}(b). In the presence of an external
electric field,
the non-equlibrium Hall currents flow inside the
white and black regions. (Relation of this ``bulk" transport description to
more conventional
Landauer-B\"uttiker formalism will be discussed later on). We assume that
temperature is so low
that scattering inside the bulk regions is absent. In this case, due to the
continuity conditions,
the currents are not allowed to cross the edges and must flow along them
focusing at the saddle
point\cite{dykhne}.
 Let
us define the currents entering and leaving the white quadrants  $I$ and $I'$
and those in the black
quadrants  $J$ and $J'$ as shown in Fig.~\ref{quantum perco}(b).
For any distribution of
the current densities across
the quadrants, one can relate the net currents to the four electric potentials
on edges $V_1$
through $V_4$, as given by

\begin{eqnarray}
I = \sigma_1 (V_2-V_1), \ \
I' = \sigma_1  (V_3-V_4);  \nonumber \\
J = \sigma_2 (V_3-V_2),\ \
J' = \sigma_2  (V_4-V_1).
\label{jv}
\end{eqnarray}
As one can easily check, current conservation is met only if $V_2-V_1=V_3-V_4$.
Thus we have in fact $I=I'$ and $J =
J'$, i.e. there is a unique white-to-white current
and a unique black-to-black one for each saddle point.

The pattern of the current distribution in the sample
is determined by the ``white" and ``black" currents $I_i$ and
$J_i$, for each vertex $i$ in the sample
(Fig.~\ref{quantum perco}(a)).  This in general
can be quite complex.  Let $\vec i$ and $\vec j$ be the average partial
current densities for the white and black phases respectively. The
total current density is written $\vec j_{tot} = \vec i + \vec j$.
The central rigorous result in this paper is the following theorem:

{\bf Theorem:} If and only if $\vec i \perp \vec j$, the effective
conductivity tensor $\hat \sigma$
obeys the semi-circle relation in Eq.~(\ref{semi circle}).
This theorem reduces the semi-circle relationship between
$\sigma_{xx}$ and $\sigma_{xy}$ to a purely
geometrical property of current distributions in the two phases.

{\bf Proof:} Let us introduce, in addition to the partial
current densities $\vec i$ and $\vec j$, the average ``white'' and ``black''
electric fields
$\vec e$ and $\vec f$, related to the former by

\begin{eqnarray}
\vec i = \sigma_1 \hat \epsilon \cdot \vec e, \ \
\vec j = \sigma_2 \hat \epsilon \cdot \vec f,
\label{ef}
\end{eqnarray}
where $\hat \epsilon$
is the matrix of clockwise rotation by $90^0$.  These relations are valid only
if scattering is
absent everywhere except at separated points (vertices). One can obtain them,
e.g., by
drawing a vertical line across the sample
as shown in Fig.~\ref{quantum perco}(a).
Adding total voltage drop along it, and the total
current across it, for white and
black regions separately, one obtains
$i_x=\sigma_1 e_y$ and $j_x=\sigma_2 f_y$.
Repeating this procedure for a horizontal line
yields the remaining relations in Eqs.~(\ref{ef}).
Now suppose $\vec i$ and $\vec j$ are perpendicular
to each other.  This  can be written as
\begin{equation}
\vec e = \beta \hat \epsilon \cdot \vec f  ,
\label{eperpf}
\end{equation}
where $\beta$ is some dimensionless parameter.
The total current density and field
are then given by
$
\vec j_{{\rm tot}} = \vec i + \vec j, \
\vec e_{{\rm tot}} = \vec e + \vec f.$
Using Eqs.~(\ref{eperpf}) and (\ref{ef}),
we can relate $\vec j_{\rm tot}$ to
$\vec e_{\rm tot}$ directly,
$\vec j_{\rm tot} = \hat \sigma_{\rm eff} \cdot \vec e_{\rm tot}$,
where the effective conductivity tensor $\hat \sigma_{{\rm eff}}$
is given by
\begin{eqnarray}
\hat \sigma_{{\rm eff}}
= \frac{\sigma_1+\sigma_2}{2}\, \hat \epsilon  +
\frac{\sigma_1-\sigma_2}{2}\,\hat \epsilon \,
\frac{\beta+\hat \epsilon}{\beta-\hat \epsilon} .
\label{sigma inv}
\end{eqnarray}
It is easy to show that Eq.~(\ref{sigma inv}) implies
\begin{equation}
{\rm det} \left[ \hat \sigma_{{\rm eff}} - \hat \epsilon
\,\frac{\sigma_1+\sigma_2}{2} \right]
= \frac{(\sigma_1-\sigma_2)^2}{4}  ,
\label{det}
\end{equation}
indepedent of the value of $\beta$.
This is equivalent to the semi-circle relation (\ref{semi circle}).
 It is also easy to see that if the condition
$\vec i \perp \vec j$ is not satisfied,
we would have $\hat \beta = \beta \hat \Gamma$,
where $\hat \Gamma$ is some rotation matrix.
In this case, the determinant in Eq.~(\ref{det}) would
depend on $\beta$.  Thus $\vec i \perp \vec j$
is also a necessary condition for the semi-circle
relation (\ref{semi circle}). Q.E.D.

We now verify analytically that $\vec i \perp \vec j$ holds for an incoherent
(classical) system at high $T$.  This regime is characterized by
a relatively short phase-coherence length, $l_{\phi} \ll \xi_p$, where
$\xi_p$ is the percolation correlation length determining the
typical distance between vertices.
In this case, scattering at a vertex has a local nature,
the black and white currents at each vertex $i$ are
uniquely related by
\begin{equation}
I_i = \alpha_i J_i  ,
\label{class}
\end{equation}
where parameter $\alpha_i$ depends on the scattering properties of
this vertex\cite{dykhne}. For our purposes, it suffices to consider this
quantity as
a phenomenological parameter which fluctuates from vertex to vertex. As one can
prove,
Eq.~(\ref{class}) and the condition that the sum
of currents entering each (black or white) region must
be zero, uniquely define all the currents
 $\{I_i , J_i\}$, given the avaerage current density $\vec j_{{\rm tot}}$.
Assuming that chirality is absent in the neighborhood of all the saddle points,
parameters
$\alpha_i$ must change sign when the magnetic field is reversed
 \begin{equation}
\alpha_i (B) = -\alpha_i(-B)  .
\label{alpha}
\end{equation}

We now show that this condition leads to $\vec i \perp \vec j$.
 Let $\{I_i , J_i\}$
be the current distribution for some magnetic field $B$ and some
$\vec j_{{\rm tot}}$, with $\vec i$ and $\vec j$ being the partial current
densities. At
the opposite direction of the magnetic field, $B' = -B$, there exists a
solution
 $\{I'_i,J'_i\}$  such that $I'_i=I_i$ and $J'_i=-J_i$. The partial current
densities are
$\vec i' = \vec i$ and $\vec j' = -\vec j$. On the other hand,
since the system is
isotropic and non-chiral at large length scales, the pair of vectors $\vec i',
\vec j'$
can be obtained from
the pair $\vec i, \vec j$ by rotation and reflection. These
conditions are compatible only if
$\vec i \perp \vec j$. The proof given can be easily generalized to the more
complex
case where each separate
vertex has a random chirality, but the
system as a whole
is non-chiral.

Before going on to show $\vec i \perp \vec j$ in the
quantum transport regime, let us discuss a rather general issue of the relation
between bulk and edge-state transport pictures when applied to our problem.
The bulk current description employed above \cite{woltjer}-\cite{cooper}
differs from
the conventional
edge-state formalism which assumes that current
flows directly on the edge.  This discrepancy
comes from the use of single-electron picture in the latter theory:
 non-equilibrium current in the bulk is impossible
since both concentration of electrons and built-in
electric field (in our case, that from random potential)
are fixed.  Adding electrons to the edge by shifting
chemical potentials results in this picture only
in an additional edge current.  But since real
electrons do have charge, accumulation of electrons
at edge-states (and of localized quasiparticles in the bulk)
generates an electric field $\vec E(\vec r)$
inside the bulk regions
in Fig.~\ref{quantum perco}(a), causing
additional Hall current $\vec j(\vec r)$ to flow in the bulk\cite{thouless}.
Using Lorentz transformation, one can show
that a homogeneous interacting electron system
subject to a homogeneous electric field
acquires a drift velocity equal to the classical value for a single electron.
In the spirit of linear response, it is natural
to assume that the same is true locally for a weakly
inhomogeneous case: $\vec j(\vec r) = \sigma_{xy}(\vec r) \hat \epsilon
\cdot \vec E(\vec r)$, where $\sigma_{xy}(\vec r)= en(\vec r)c/B$.
 Our next
assumption is that quasi-particles in the bulk
are localized by short-range impurities, so that the effective concentration
$n$ which contribute to the Hall current is given by that of the incompressible
liquid.
As a result, integrating electric field and the current
density along any curve connecting the opposite
sides of an angle in Fig.~\ref{quantum perco}(b)
gives relations of Eqs.~(\ref{jv}).

Although the above arguments suggest that the local
$\sigma_{xy}$ picture employed here is more appropriate
for describing current distributions in the case of long-range electron
interaction
than the conventional edge-state formalism, it
turns out that, in fact, both descriptions will yield
identical results, in the absence of scattering in the bulk.
This is because, in the edge-state formalism, the current
density can be written as

\begin{equation}
\vec j_{{\rm edge}} = \varphi(\vec r)\hat\epsilon\cdot\nabla \sigma_{xy},
\label{edge}
\end{equation}
where $e \varphi(\vec r)$ is the chemical potential function.
In the local $\sigma_{xy}$ picture, the corresponding
current density is given by
\begin{equation}
\vec j_{{\rm loc}} = -\sigma_{xy}\hat\epsilon\cdot\nabla \varphi(\vec r),
\label{local}
\end{equation}
where $\varphi(\vec r)$ is now the electric potential function.
For step-like distribution of $\sigma_{xy}$ [from
white region to black region in Fig.~\ref{quantum perco}(a)],
Eq.~(\ref{edge}) yields a delta-function on the edge
with an integrated current $(\sigma_2-\sigma_1) \varphi_{{\rm edge}}$.
As is easy to check, the difference between
the two current densities (\ref{edge}), (\ref{local}) is a curl which yields
zero
upon integration along any line throughout the whole sample,
when the two $\varphi$ functions
in the two formalisms are identified to one another.
Thus we are free to choose either formalism according to
our convenience in our problem.

We now proceed to show that $\vec i \perp \vec j$
for coherent quantum percolation systems.
At very low temperatures,
different clusters of the same liquid couple to each
other through quantum tunneling to form
a coherent wavefunction whose size is larger
than the classical cluster size $\xi _p$ and diverges
when approaching the percolation threshold.
Quantum coherence implies that currents at a given saddle
point can depend on those at other ones, so that the
local relation Eq.~(\ref{class}) is no longer valid.
Moreover, it is no longer sufficient to use currents $\{I_i,J_i\}$
to describe a given state of the system, since complex quantum amplitudes
must be used instead.  Thus we need to resort to
a particular quantum percolation model.
Our choice is the fully quantum coherent network model first suggested by
Chalker and  Coddington\cite{chalker-coddington}, which has proven quite
successful
in the study of localization length exponent in quantum Hall
systems\cite{chalker-coddington},\cite{kivelson}. In this model, a complex
amplitude $A_k$ is
assigned to each  bond of an oriented regular square network [see
Fig.~\ref{quantum perco}(c)].
At the vertex marked $\Omega$ in Fig.~\ref{quantum perco}(c), the four quantum
amplitudes are related by a scattering matrix given by
\begin{equation}
\left({{A_4}\atop {A_3}}\right)=
\left({{e^{i\phi_4}}\atop 0}{0\atop {e^{i\phi_3}}}\right)
\!\left(
{{\cosh \gamma} \atop{\sinh \gamma}}
{{\sinh \gamma} \atop{\cosh \gamma}}\right)
\!\left({{e^{i\phi_1}}\atop 0}{0\atop {e^{i\phi_2}}}\right)
\!\left({{A_1}\atop {A_2}}\right)
\label{scatt}
\end{equation}
where $\phi_k$ are random phase factors
which model the effect of random path lengths
in an actual random geometry such as
in Fig.~\ref{quantum perco}(a), $\gamma$ is the tunneling parameter which, for
the sake of
simplicity, is set to be the same for all the vertices. For vertices oriented
like the one
marked $\Gamma$  in Fig.~\ref{quantum perco}(c), the bonds denoted
$A_1,...,A_4$ should be rotated by
$90^0$ with respect to those with orientation $\Omega$.
The limits $\gamma\to 0$ and $\gamma\to \infty$ correspond to
$100\%$ percolation of the white or
the black regions, respectively. The value $\gamma_c = \ln (1+\sqrt{2})$
defines the quantum percolation threshold\cite{chalker-coddington}. The system
of Eqs.~(\ref{scatt})
written for all the saddle points determines the distribution of the quantum
amplitudes,
i.e. the wavefunction.

In order to complete the construction of our formalism, we need to adapt this
model
to the study of transport properties.
Original quantum percolation model\cite{chalker-coddington} used the
edge-transport theory as a
starting point, its classical analog being the ``network model'' suggested in
Ref. \cite{kucera}.
The squares of amplitudes $|A_k|^2$ were identified with the edge currents
entering and leaving the
vertices as shown in Fig.~\ref{quantum perco} (c).
Unitarity of the scattering matrix in
Eq.~(\ref{scatt}) takes care of current conservation:
$|A_1|^2-|A_2|^2=|A_4|^2-|A_3|^2$. Obviously,
the standard edge-transport theory cannot be applied consistently to the
quantum case since the
chemical potential it deals with cannot be introduced for separate bonds
belonging to the same
wavefunction. Hence we identify $|A_k|^2$ with the fictious edge current
$(\sigma
_2-\sigma _1)V_k$ discussed above in a formal
sense, with
$V_k$ being the {\it electric} potential of the bond. Then, as one
can see from Eqs.~(\ref{jv}), the
physical currents in the white and black quadrants [Fig.~\ref{quantum
perco}(b)]
are related to the
differences in the edge currents on the adjoining edges, given by

\begin{equation}
I= \sigma_1 \left(|A_2|^2-|A_1|^2\right),\ \ J = \sigma_2
\left(|A_3|^2-|A_2|^2\right), \label{IJtotal}
\end{equation}
apart from a constant factor
$\sigma_1-\sigma_2$. Since we study small non-equilibrium currents
everywhere in this work,
Eqs.~(\ref{IJtotal}) can be linearized

\begin{eqnarray}
I = \sigma_1 [Re(A_2 a^*_2)-Re(A_1 a^*_1)], \nonumber \\
J = \sigma_2 [Re(A_3 a^*_3)-Re(A_2 a^*_2)],
\label{IJ}
\end{eqnarray}
where $a_k$ are small perturbations of the quantum amplitudes due to an
applied external voltage $V$,
while $A_k$ are the equilibrium values in the absense of $V$. Quantities
$Re(A_ka_k^*)$ are the
non-equilibrium electric potentials.

The remaining task is to solve the system of linear Equations (\ref{scatt})
numerically with proper
boundary conditions and find the average black and white current densities
using Eqs.~(\ref{IJ}).
We chose standard periodic boundary conditions
on the left and right boundaries of a rectangular sample in Fig.~\ref{quantum
perco}(c). Since Eqs.~(\ref{scatt}) are linear, they can be solved separately
for $A_k$
and $a_k$, but with different boundary conditions. We let all equilibrium
amplitudes $A_k$ on the
incoming bonds on the upper and lower boundaries to be unity. For the same
sample, we let incoming
amplitudes $a_k$ on the lower boundary to be $+V/2$ and those on the upper
boundary to be $-V/2$.
This simulates
the situation where a small two-terminal voltage $V$ is applied across the
Corbino disc-shaped sample attached to two metallic contacts. Indeed,
as shown by one of us when
studying the role of contacts in the non-uniform Hall conductors\cite{ruzin}, a
potential step near
the contact is possible if and only if local $\sigma_{xy}$ increases to the
right when looking into
the sample from the contact. This is the case for the bonds arrowed as
outcoming ones in
Fig.~\ref{quantum perco}(c) ($\sigma_1 < \sigma_2$ is implied). The incoming
bonds,
on the contrary,
must have potentials of the contacts. This is completely analogous to the
``ballistic" boundary conditions in the conventional edge-state transport
theory where
the incoming (leaving a contact) one-electron states are assumed to have the
chemical potential of the contact.

For each realization of $\{ \phi_k\}$,
we can extract the currents densities $\vec i$ and $\vec j$ by
adding up the corresponding currents at the vertices, according to
Fig.~\ref{quantum perco}(c). Namely,
$i_x$ and $j_y$ are found by averaging
$I$ and $J$ currents, respectively, over all the vertices with orientation
$\Omega$
(Fig.~\ref{quantum perco}(c)).
Similarly, $i_y$ and $j_x$ are extracted from the vertices $\Gamma$. The total
number of rows of vertices in either direction is even. The system size $L_x
\times L_y$
ranges from $20\times 20$ to $50\times 82$, and we averaged over 2,000
realizations to find
the ensemble averaged $\langle \vec i\rangle $ and $\langle \vec j
\rangle$ currrents.  We note the importance of ensemble
averaging: in the fully quantum mechanical
case studied here, there is no simple self-averaging even in a large
sample.
Self-averaging occurs, in a physical system,
only on the length scale of an inelastic phase-breaking length,
which at low temperatures can be very large, the effect of which is replaced
in our simulation by a finite system size. Our result for $\gamma=\gamma_c =
\ln (1+\sqrt{2})$ can be summarized
as
\begin{equation}
\cos \alpha = \frac{\langle \vec i\rangle \cdot \langle \vec j \rangle}
{|\langle \vec i\rangle| |\langle \vec i\rangle|} = 0.01 \pm 0.02 .
\end{equation}
This shows that the white and black currents
are indeed perpendicular to each other to a high degree of accuracy.  We have
checked that this null value for $\cos \alpha$ persists when the size of the
sample is changed. Neither did we find any dependence of $\cos \alpha$ on
the sample shape for any $L_y/L_x >1.3$.
Finally, we made sure that $\cos\alpha$, in such samples, is insensitive
 to changes of $\gamma$ around
the critical value $\gamma_c$. The interval $\gamma$ in which the ratio of
$|\vec i|/|\vec j|$ changes by 2 orders of magnitude was studied.
(A weak shape dependence at $\gamma = \gamma_c$ was revealed in short samples:
at $L_y/L_x =0.3$,
$\cos\alpha = 0.14\pm 0.02$. We interpret this as
an artifact of closely placed metallic contacts;
since when
the localization length becomes smaller than the
sample size as we take $\gamma \not= \gamma_c$, $\cos \alpha$ restores its null
value.) In
combination with the Theorem discussed before, we have thus successfully
demonstrated that the
universal semi-circle relation between $\sigma_{xx}$ and $\sigma_{xy}$ holds in
the
regime of quantum transport.

We thank F. Zeng for assistance in simulations,
and S. Kivelson for many stimulating discussions.
This work was supported in part by the ONR
Grant N00014-90-J-1829, and the DOE Grant DE-FG03-88ER45378.

\begin{figure}
\caption{(a) Schematic illustration of the random quantum percolation
model for describing the transition between successive
quantum Hall plateaus.  (b) A single saddle point magnified.
(c) The Chalker-Coddington network model ($\sigma_1 < \sigma_2$).}
\label{quantum perco}
\end{figure}

\end{document}